\documentclass[aps,nofootinbib,12pt]{revtex4}
\usepackage{amsmath}
\usepackage{amsfonts}
\usepackage{amssymb}
\usepackage{color}
\usepackage{graphicx}
\usepackage{mathrsfs}
\usepackage{slashbox}

\def\bc{\begin{center}}
\def\nno{\nonumber}
\def\ec{\end{center}}
\def\be{\begin{eqnarray}}
\def\ee{\end{eqnarray}}

\definecolor{dyellow}{rgb}{1.,0.8,.0}
\definecolor{myblue}{rgb}{.1,.1,.7}
\definecolor{dcyan}{rgb}{.0,.6,.6}
\definecolor{dmagenta}{rgb}{0.6,0.0,0.6}
\definecolor{brown}{rgb}{0.6,0.2,0.}
\definecolor{darkblue}{rgb}{.0,.0,0.5}
\definecolor{darkred}{rgb}{0.75,0.0,0.0}
\definecolor{orange}{rgb}{1.,.6,.0}
\definecolor{dorange}{rgb}{0.8,.4,.0}
\definecolor{darkgreen}{rgb}{0.0,0.6,0.0}
\definecolor{purple}{rgb}{.4,.0,.4}
\definecolor{lightgrey}{rgb}{0.7, 0.7, 0.7}
\definecolor{grey}{rgb}{0.4, 0.4, 0.4}
\def\al{\alpha}
\def\ga{\gamma}
\def\dl{\delta}
\def\eps{\epsilon}
\def\ka{\kappa}
\def\la{\lambda}

\def\si{\sigma}

\def\pa{\partial}

\def\Ga{\Gamma}
\def\Dl{\Delta}

\def\Om{\Omega}
\def\h{\hat}

\begin{document}%\large\bf

\title{Holographic phase transitions of p-wave superconductors in Gauss-Bonnet gravity with back-reaction} \vskip 2cm \vskip 2cm
\author{Rong-Gen Cai}\email{cairg@itp.ac.cn}
\author{Zhang-Yu Nie}\email{niezy@itp.ac.cn}
\author{Hai-Qing Zhang}\email{hqzhang@itp.ac.cn}
\address{Key Laboratory of Frontiers in Theoretical Physics,
Institute of Theoretical Physics, Chinese Academy of Sciences,
   P.O. Box 2735, Beijing 100190, China}

%\date{November 2009}

\begin{abstract} \vspace{5mm}
We investigate the phase transitions of holographic p-wave
superconductors in $(4+1)$-dimensional
Einstein-Yang-Mills-Gauss-Bonnet theories, in a grand canonical
ensemble. Turning on the back-reaction of the Yang-Mills field, it
is found that the condensations of vector order parameter become
harder if the Gauss-Bonnet coefficient grows up or the back-reaction
becomes stronger. In particular,  the vector order parameter
exhibits the features of first order and second order phase
transitions, while only the second order phase transition is
observed in the probe limit. We discuss the roles that the
Gauss-Bonnet term and the back-reaction play in changing the order
of phase transition.

\end{abstract}
\maketitle

\section{\bf Introduction }
The AdS/CFT correspondence
\cite{Maldacena:1997re,Gubser:1998bc,Witten:1998qj,Aharony:1999ti}
provides a novel approach to study the strongly coupling systems at
finite density. Therefore, it may have some useful applications in
condensed matter physics. It has been applied to study the
holographic shear viscosity
\cite{Policastro:2001yc,Kovtun:2003wp,Buchel:2003tz,Kovtun:2004de},
holographic superconductors \cite{Gubser:2008px,Hartnoll:2008vx} and
holographic (non)fermi-liquids
\cite{Lee:2008xf,Liu:2009dm,Cubrovic:2009ye}. In this paper, we will
focus on the holographic p-wave superconductors
\cite{Gubser:2008wv,Basu:2009vv,Ammon:2009xh}.

The phenomena of superconducting can be explained by the
spontaneously breaking of U(1) gauge symmetry
\cite{Weinberg:1986cq}. In the p-wave superconductor, the
rotational symmetry is also broken by a special direction of some
vector field in addition. This could be achieved by the
condensation of a charged vector field. And the holographic
modeling of this picture could be simply realized by adding an
SU(2) Yang-Mills field to an AdS black hole
background\cite{Gubser:2008wv}. In this case, after making some
ansatz of the SU(2) field, one U(1) subgroup of SU(2) is
considered as the electromagnetic gauge group, in addition, a
gauge boson generated by another SU(2) generator is charged under
this U(1) subgroup through the nonlinear coupling of the
non-Abellian gauge fields. In this setup, the superconductor phase
transition is studied, and conductivities show some anisotropic
behavior. Further more, the holographic p-wave superconductor with
back-reaction was investigated in Ref.~\cite{Ammon:2009xh}. The
crucial point is that the back-reaction will dramatically change
the order of the phase transition. More specifically, when the
matter field coupling goes beyond a critical value, the former
second order phase transition with a $1/2$ mean-field theory
critical exponent near the critical temperature
\cite{Gubser:2008wv,Ammon:2009xh,Cai:2010cv,
Barclay:2010up,Gregory:2010yr,Barclay:2010nm} will be changed to a
first order phase transition.

In a previous paper \cite{Cai:2010cv}, we studied the holographic
Gauss-Bonnet p-wave superconductors in the probe limit. The
holographic Gauss-Bonnet superconductors are also discussed in
\cite{Gregory:2009fj,Pan:2009xa,Brihaye:2010mr,Pan:2010at,Siani:2010uw,Jing:2010cx}.
In this paper we will investigate the holographic p-wave
superconductors with back-reaction in order to find out how the
matter couplings and the Gauss-Bonnet coefficient  affect the phase
transition of the p-wave superconductor. We find that the bigger the
Gauss-Bonnet coupling is, the bigger the condensation value of the
order parameter is, and the lower the critical temperature is. This
reflects that the big Gauss-Bonnet coefficient will make the
superconducting phase transition hard, which is consistent with our
previous conclusions \cite{Cai:2010cv}. Besides, we also find that
the stronger the matter field couples to the background, the harder
the condensation to be formed. In addition, we find that the phase
transition will change from second order to first order when the
back-reaction is strong, which is similar to the discussions of
\cite{Ammon:2009xh}. In  grand canonical ensemble, we study the free
energy and entropy of the p-wave superconductor which also support
our claim of the change of phase transitions.

This paper is organized as follows: We will set up our model of
the holographic superconductors in Sec.~\ref{sect:setup} and study
the condensation behavior of the vector order parameter for
different Gauss-Bonnet coefficients and different matter field
couplings in Sec.~\ref{sect:solutions}. In
Sec.~\ref{sect:thermodynamics}, we study the thermodynamics of the
p-wave superconductor by exploring the free energy and entropy. We
draw our conclusions in Sec.~\ref{sect:conclusion}.

\section{Holographic set up of p-wave superconductors}
\label{sect:setup}

We consider the Einstein-Gauss-Bonnet gravity with an SU(2)
Yang-Mills field in $(4+1)$-dimensional asymptotically AdS
space-time. The action is
 \begin{equation}\label{action}
S=\int d^5 x \sqrt{-g} \Big[\frac{1}{2
\kappa_5^2}\Big(R+\frac{12}{L^2}+\frac{\alpha}{{2}} (R^2-4R^{\mu\nu}
R_{\mu\nu}+R^{\mu\nu\rho\sigma}R_{\mu\nu\rho\sigma})\Big){-}\frac{1}{4
\hat{g}^2}\Big(F^a_{\mu\nu}F^{a\mu\nu}\Big)\Big]+S_{{bdy.}}
\end{equation}
where $\kappa_5$ is the five dimensional gravitational constant with
$2\ka_5^2=16\pi G_5$, and $G_5$ a $(4+1)$-dimensional Newton
gravitational constant, $\hat{g}$ is the Yang-Mills coupling
constant and $L$ is the AdS radius. The SU(2) Yang-Mills field
strength is
 \be F^a_{\mu\nu}=\partial_\mu A^a_\nu-\partial_\nu
A^a_\mu + \epsilon^{abc}A^b_\mu A^c_\nu \ee
 where $a,b,c=(1,2,3)$ are the indices of the generators of SU(2)
 algebra. $\mu,\nu=(t,r,x,y,z)$ are the labels of space-time with
 $r$ being the radial coordinate of AdS. The $A^a_{\mu}$ are the
 components of the mixed-valued gauge
 fields
 $A=A^a_{\mu}\tau^adx^{\mu}$, where $\tau^a$ are the SU(2)
 generators with commutation
relation $[\tau^a,\tau^b]=\eps^{abc}\tau^c$.
  $\epsilon^{abc}$ is the totally antisymmetric tensor
with $\epsilon^{123}=+1$. The quadratic curvature term is the
Gauss-Bonnet term with $\alpha$ the Gauss-Bonnet coefficient and $
R^{\mu}_{\ \nu\rho\si}=\pa_{\rho}\Ga^{\mu}_{\ \nu\si}-\cdots $.
$S_{bdy}$ includes boundary terms that do not affect the equations
of motion, namely the Gibbons-Hawking surface term, as well as
counter-terms required for the on-shell action to be finite. We will
write $S_{bdy}$ term explicitly in Sec.~{\ref{sect:thermodynamics}}.

The Einstein field equations can be derived from the above action as
% \be R_{\mu\nu}-\frac1 2
% g_{\mu\nu}(R+\frac{12}{L^2})+\frac{\al}{2}[H_{\mu\nu}-\frac1
 %2g_{\mu\nu}(R^2-4R^{\rho\si}R_{\rho\si}+R^{\al\beta\tau\rho}R_{\al\beta\tau\rho})]=\kappa_5^2T_{\mu\nu}\ee}
% or,
 \be R_{\mu\nu}-\frac1 2
 g_{\mu\nu}(R+\frac{12}{L^2})+\frac{\al}{2}[H_{\mu\nu}-\frac1
 2g_{\mu\nu}\frac H 2 ]=\kappa_5^2T_{\mu\nu}\ee
with \be T_{\mu\nu}&=&\frac{1}{\h g^2}
\text{tr}(F^{a}_{\rho\mu}F^{a\rho}_{\ \
\nu}-\frac1 4g_{\mu\nu}F^a_{\rho\si}F^{a\rho\si}),\\
   H_{\mu\nu}&=&2RR_{\mu\nu}+2R_{\mu\al\tau\rho}R_{\nu}^{\
  \al\tau\rho}-4R_{\mu\rho}R^{\rho}_{\ \nu}+4R^{\si}_{\
  \rho}R^{\rho}_{\ \nu\mu\si},\\
  H&=&H^{\mu}_{\ \mu}.\ee
  where ``tr" takes the trace over the indices of SU(2) generators.
The Yang-Mills equations of motion are:
 \be \nabla_{\mu}F^{a\mu\nu}=-\eps^{abc}A^b_{\mu}F^{c\mu\nu}.\ee

 Following  Refs.~\cite{Gubser:2008wv,Basu:2009vv,Ammon:2009xh}, we choose the ansatz of the gauge fields as
  \be\label{ansatz} A(r)=\phi(r)\tau^3dt+w(r)\tau^1dx.\ee
 In this ansatz we regard the U(1) symmetry generated by $\tau^3$ as the U(1) subgroup of
 SU(2). We call this U(1) subgroup as U(1)$_3$. The
gauge boson with nonzero component $w(r)$ along $x$ direction is
charged under $A^3_t=\phi(r)$. According to AdS/CFT dictionary,
$\phi(r)$ is dual to the chemical potential in the boundary field
theory while $w(r)$ is dual to the $x$ component of some charged
vector operator $J$. The condensation of $w(r)$ will spontaneously
break the U(1)$_3$ gauge symmetry and induce the phenomena of
superconducting on the boundary field theory.

Our metric ansatz following
 Ref.~\cite{Ammon:2009xh,Manvelyan:2008sv} is
  \be\label{ansatzg}
  ds^2=-N(r)\si(r)^2dt^2+\frac{1}{N(r)}dr^2+r^2f(r)^{-4}dx^2+r^2f(r)^2(dy^2+dz^2).\ee
with
  \be N(r)=\frac{r^2}{2\al}\Big(1-\sqrt{1-\frac{4\al}{L^2}+\frac{4\al
  m(r)}{r^4}}\Big).\ee
  where $m(r)$ is a function related to the mass and charge of the
  black hole. The reason for this metric ansatz is that: the
  back-reaction of nonzero $w(r)$ will change the background of
  space-time. The condensation of $w(r)$ will preserves only SO(2)
  symmetry of the spatial direction, {\it i.e.}, $(y,z)$-direction.
  The horizon of the black hole is located at $r_h$ while the
  boundary of the bulk is at $r_{bdy}\rightarrow \infty$. Note
  that when $r\rightarrow \infty$,
  \be N(r)\sim\frac{r^2}{2\al}\Big(1-\sqrt{1-\frac{4\al}{L^2}}\Big).\ee
  So we can define an effective radius $L_c$ of AdS space-time as
   \be L_c\equiv L\sqrt{\frac{1+U}{2}},\quad
 U=\sqrt{1-\frac{4\al}{L^2}}.\ee
 From this relation we can see that in order to have a well-defined
 vacuum for the gravity theory, there is an upper bound for $\al\leq
 L^2/4$. The saturation $\al=L^2/4$ is called Chern-Simons limit. If
 we further consider the causality constraint of the boundary CFT,
 there is an additional constraint on the Gauss-Bonnet coefficient
 with $-7L^2/36\leq\al\leq9L^2/100$ in five dimensions~\cite{Brigante:2007nu,
 Brigante:2008gz,Buchel:2009tt,
Hofman:2009ug,deBoer:2009pn,Camanho:2009vw,Buchel:2009sk}.

 The Hawking temperature of this black hole is
 \be \label{temp}T=\frac{\si N'}{4 \pi}\Big|_{r=r_h}=\Big(\frac{\si}{\pi
 L^2}-\kappa_g^2\frac{\phi'^2}{12\pi \si}\Big)r \Big|_{r=r_h}\ee
where, $\kappa_g\equiv\kappa_5/\hat{g}$  is regarded as the
effective matter field coupling, and `` $'$ " denotes the derivative
with respect to $r$. The Bekenstein-Hawking entropy of the black
hole is :
 \be S=\frac{A}{4G_5}=\frac{2\pi A}{\ka_5^2}=\frac{2\pi}{\ka_5^2}V
 r_h^3,\ee
 where $A$ denotes the area of the horizon and $V=\int d^3x$.

 The Einstein and Yang-Mills equations of motion
 with the ansatz (\ref{ansatz},\ref{ansatzg})
 can be explicitly written as
 \be\label{Neq}
 N'&=&-\frac{\ka_g^2 r f^4w^2\phi^2}{3r^2N\si^2-6\al
 N^2\si^2(rf^{-2})'(rf)'^2}-\frac{\ka_g^2r^3\phi'^2}{3r^2\si^2-6\al
 N\si^2(rf^{-2})'(rf)'^2}\nno\\&&-\frac{r(NL^2(6f^2+6r^2f'^2+\ka_g^2f^6w'^2)-12f^2r^2)
 +24L^2r\al N^2f'(rf)'(f^{-1}rf')'}{3L^2r^2f^2-6L^2\al
 f^2N(rf^{-2})'(rf)'^2},\\
 \nno\\
 \label{sieq}
 \si'&=&\frac{\ka_g^2rf^4w^2\phi^2}{6r^2N^2\si-12\al
 N^3\si(rf^{-2})'(rf)'^2}-\frac{\ka_g^2r^3\phi'^2}{6r^2N\si-12\al
 N^2\si(rf^{-2})'(rf)'^2}\nno\\
 &&+\frac{12 r^3 \sigma  f^2+L^2 \sigma
   \left(\ka_g^2 r N w'^2 f^6-3f^2
   (N' r^2+2 N (r-\alpha
   (rf^{-2})' (r f)'^2
   N'))+6 r^3 N
   f'^2\right)}{6L^2fN(r^2-2\al
 N(rf^{-2})'(rf)'^2)},\nno\\\\
 \label{feq}
 f''&=&f_1+f_2+f_3,\\
 \label{phieq}
 \phi''&=&\frac{f^4w^2\phi}{r^2N}+(-\frac3
 r+\frac{\si'}{\si})\phi',\\
 \label{weq}
 w''&=&-\frac{w\phi^2}{N^2\si^2}-w'(\frac1
 r+4\frac{f'}{f}+\frac{N'}{N}+\frac{\si'}{\si}),\ee

where
 \be
 f_1&=&\bigg(-\ka_g^2r^2f^7w^2\phi^2+\ka_g^2\al f^5 N
 w^2\phi^2(2f-rf')(rf)'\bigg)\bigg/\bigg(3r^4f^2N^2\si^2\nno\\&&+3\al r^2N^2\si
 (2r^2N\si
 f'^2-2r^2ff'(\si N'+2N\si')-f^2(r\si
 N'+2N(r\si)'))\nno\\&&+6\al^2rN^3\si(rf)'^2(\si N'+2N\si')\bigg)\\
 f_2&=&\bigg(\ka_g^2r^2f^7\si w'^2-\ka_g^2\al
 f^5N\si(2f-rf')(rf)'w'^2\bigg)\bigg/\bigg(3r^4f^2\si\nno\\&&+3\al
 r^2(2r^2N\si f'^2-2r^2ff'(\si N'+2N\si')-f^2(r\si
 N'+2N(r\si)'))\nno\\&&+6\al^2rN(rf)'^2(\si N'+2N\si')\bigg)\\
f_3&=&\bigg(-L^2 r^3 \sigma  f' \left(f \left(r \sigma
   N'+N \left(3 \sigma +r \sigma
   '\right)\right)-r N \sigma  f'\right)
   f^3\nno\\&&+r \alpha f f' \left[\left(L^2 \sigma ^2
   N'^2 r^2+N \left(\ka_g^2 \phi '^2 L^2+2
   \sigma  N' \sigma ' L^2+12 \sigma ^2\right)
   r^2+2 L^2 N^2 \sigma  \left(2 \sigma +r \sigma
   '\right)\right) f^3\right.\nno\\&&\left.+r f' \left(-4 L^2 N^2
   \sigma ^2+L^2 r^2 N'^2 \sigma ^2+r N
   \left(\ka_g^2 r \phi '^2 L^2+2 r \sigma  N'
   \sigma ' L^2+2 \sigma ^2 \left(6 r-L^2
   N'\right)\right)\right) f^2\right.\nno\\&&\left.-2 L^2 r^2 N \sigma
    f'^2 \left(2 r \sigma  N'+3 N f\left(2
   \sigma +r \sigma '\right)\right) -4 L^2 r^3
   N^2 \sigma ^2 f'^3\right] \nno\\&&+4 L^2 \alpha ^2
   N^2 \sigma  f' \left(f+r f'\right)^2
   \left(-f^2+r f' f+r^2 f'^2\right)
   \left(\sigma  N'+2 N \sigma '\right)\bigg)\bigg/\nno\\&&\bigg(L^2 r^4 N \sigma ^2 f^4+2 L^2 r \alpha ^2 N^2
   \sigma  \left(f+r f'\right)^2 \left(\sigma
   N'+2 N \sigma '\right) f^2\nno\\&&+L^2 r^2 \alpha
   N \sigma  \left(2 N \sigma  f'^2 r^2-2
   f f' \left(\sigma  N'+2 N \sigma
   '\right) r^2-f^2 \left(r \sigma  N'+2 N
   \left(\sigma +r \sigma '\right)\right)\right)
   f^2\bigg).  \ee

There are four useful scaling symmetries in the above equations:

\be \label{scaling1}  &(I)   &\quad f\rightarrow\la f,\quad
 w\rightarrow\la^{-2}w,\\
 \label{scaling2}
 &(II)  &\quad\si\rightarrow \la\si,\quad \phi\rightarrow\la\phi,\\
 \label{scaling3}
 &(III) &\quad r\rightarrow \la r, \quad m\rightarrow\la ^4 m,
 \quad \omega \rightarrow\la \omega,\quad
 \phi\rightarrow\la \phi,\quad N\rightarrow\la ^2 N\\
 \label{scaling4}
 &(IV)  &\quad r\rightarrow \la r, \quad m\rightarrow\la ^2 m,
 \quad L \rightarrow\la L,\quad
 \phi\rightarrow {\la^{-1}\phi},\quad
 \kappa_g\rightarrow\la \kappa_g, \quad \alpha\rightarrow \la^2 \alpha.
\ee

 We can use the symmetries (\ref{scaling3},\ref{scaling4}) to set
 $r_h=1$ and $L=1$, and use symmetries (\ref{scaling1},\ref{scaling2})
 to set $\si(r\rightarrow\infty)=f(r\rightarrow\infty)=1$ in order to make the solution
 asymptotically approach to an AdS.

 \section{Two branches of solutions}
 \label{sect:solutions}
 \subsection{Analytic charged GB-AdS solution with {\it $w(r)=0$}}
 In the ansatz (\ref{ansatz}) with $w(r)=0$, there is an straightforward analytic black hole solution which
 is a charged generalization of GB-AdS black holes~\cite{Cai:2001dz,CNO}. In this case,
  \be f(r)=\si(r)=1,\quad
 \phi=\mu-\frac{Q}{2r^2},\ee
  and
 \be\label{rnads}
 N(r)=\frac{r^2}{2\al}\bigg[1-\sqrt{1+4\al(\frac{\tilde{M}}{r^4}
 -\frac{1}{L^2}-\ka_g^2\frac{Q^2}{6\ r^6})}\ \bigg],\ee
where $\tilde{M}$ is related to the mass of the black hole, $Q$ is
the charge of the black hole and $Q=2\mu r_h^2$. Using the formula
(\ref{temp}), the temperature for this charged GB-AdS black hole is
 \be T=\frac{r_h}{\pi L^2}-\ka_g^2\frac{ \mu^2}{3\pi r_h}.\ee

\subsection{Superconducting solutions with {\it $w(r)\neq0$}}

In order to investigate the superconducting solutions with
$w(r)\neq0$, we  have to solve numerically the set of equations
(\ref{Neq})(\ref{sieq})(\ref{feq})(\ref{phieq}) and (\ref{weq}).

First of all, we should impose  boundary conditions on the fields.
The horizon is located at $r=r_h$ with $N(r_h)=0$. On the horizon,
we should impose $\phi(r_h)=0$ for the U(1)$_3$ gauge field to have
a finite norm, and $\si(r),f(r),w(r)$ should be finite. We can
expand the fields in the powers of $(1-r_h/r)$ near the horizon,
they are
     \be\label{horizon}\text{Near Horizon}\left\{
               \begin{array}{ll}\vspace{2mm}
                  {N}&=0\Rightarrow
                  m=r_h^4/L^2+m_H^{(1)}(1-r_h/r)+\cdots\\\vspace{2mm}
                 {\si}&=\si_H^{(0)}+\si_H^{(1)}(1-r_h/r)+\cdots \\\vspace{2mm}
                 {f}&=f_H^{(0)}+f_H^{(1)}(1-r_h/r)+\cdots\\\vspace{2mm}
                 \phi&=\phi_H^{(1)}(1-r_h/r)+\phi_H^{(2)}(1-r_h/r)^2+\cdots\\\vspace{2mm}
                 w&=w_H^{(0)}+w_H^{(1)}(1-r_h/r)+\cdots ,
               \end{array}
             \right. \ee
where all the coefficients of the expansions are constants.

At the boundary $r\to \infty$, the asymptotical behavior of these
fields are
    \be \label{boundary}\text{Near Boundary}\left\{
               \begin{array}{ll}\vspace{2mm}
                  m&=m_B^{(0)}+m_B^{(2)}/r^2+\cdots\\\vspace{2mm}
                 {\si}&=\si_B^{(0)}+\si_B^{(4)}/r^4+\cdots \\\vspace{2mm}
                 {f}&=f_B^{(0)}+f_B^{(4)}/r^4+\cdots\\\vspace{2mm}
                 \phi&=\phi_B^{(0)}+\phi_B^{(2)}/r^2+\cdots\\\vspace{2mm}
                 w&=w_B^{(0)}+w_B^{(2)}/r^2+\cdots .
               \end{array}
             \right.\ee
From the AdS/CFT dictionary we know that $\phi_B^{(0)}=\mu,
\phi_B^{(2)}=\rho$, where $\mu$ and $\rho$ are respectively the
chemical potential and density of the charge at the boundary;
$w_B^{(0)}$ is the source of the boundary operator $J$ while
$w_B^{(2)}$ is the expectation value of $J$ and the nonzero
$w_B^{(2)}$ will induce superconducting phase as we have mentioned.

Near horizon (\ref{horizon}), $m_H^{(1)}, \si_H^{(1)}, f_H^{(1)},
\phi_H^{(2)}$ and $w_H^{(1)}$ can be evaluated as functions of $
\si_H^{(0)}, f_H^{(0)}, \phi_H^{(1)}$ and $w_H^{(0)}$ by
substituting the expansions into the equations of motion. Therefore,
there are four independent initial values left to be specified, {\it
i.e.}, $\si_H^{(0)}, f_H^{(0)}, \phi_H^{(1)},w_H^{(0)}$.

For the boundary conditions at infinity, we impose $\si(r)=f(r)=1$
to have an asymptotic AdS boundary, this could be reached by using
the scaling symmtries (\ref{scaling1}) and (\ref{scaling2}). We also
impose $w_B^{(0)}=0$ to turn off the the source of $J$. In the
numerical calculations, we will set $r_h=L=1$ by using the scaling
symmetries (\ref{scaling3}) and (\ref{scaling4}).

 Armed with these
equations of motion and boundary conditions we can numerically solve
the set of equations by the shooting method. Because we will work in
the grand canonical ensemble, we can fix the chemical potential
value $\mu$ and then vary the four independent near-horizon
coefficients $(\si_H^{(0)}, f_H^{(0)}, \phi_H^{(1)},w_H^{(0)})$
until we find a solution which produces the desired value of $\mu$
and $\si_B^{(0)}=f_B^{(0)}=1, w_B^{(0)}=0$.

In the following, we will present our numerical results of the
condensation value of the order parameter $J$. We will scan through
value of $\al$ from $\al=-0.19$ to $\al=0.09$ as well as $\ka_g$
from $\ka_g=0.0001$ to $\ka_g=0.45$. From the AdS/CFT
dictionary~\cite{Aharony:1999ti}, we know that the conformal
dimension of vector field in five dimension is $\la=3$, therefore,
$J^{1/3}/T_c$ is the right dimensionless quantity.

\begin{figure}
\includegraphics[width=8.8cm] {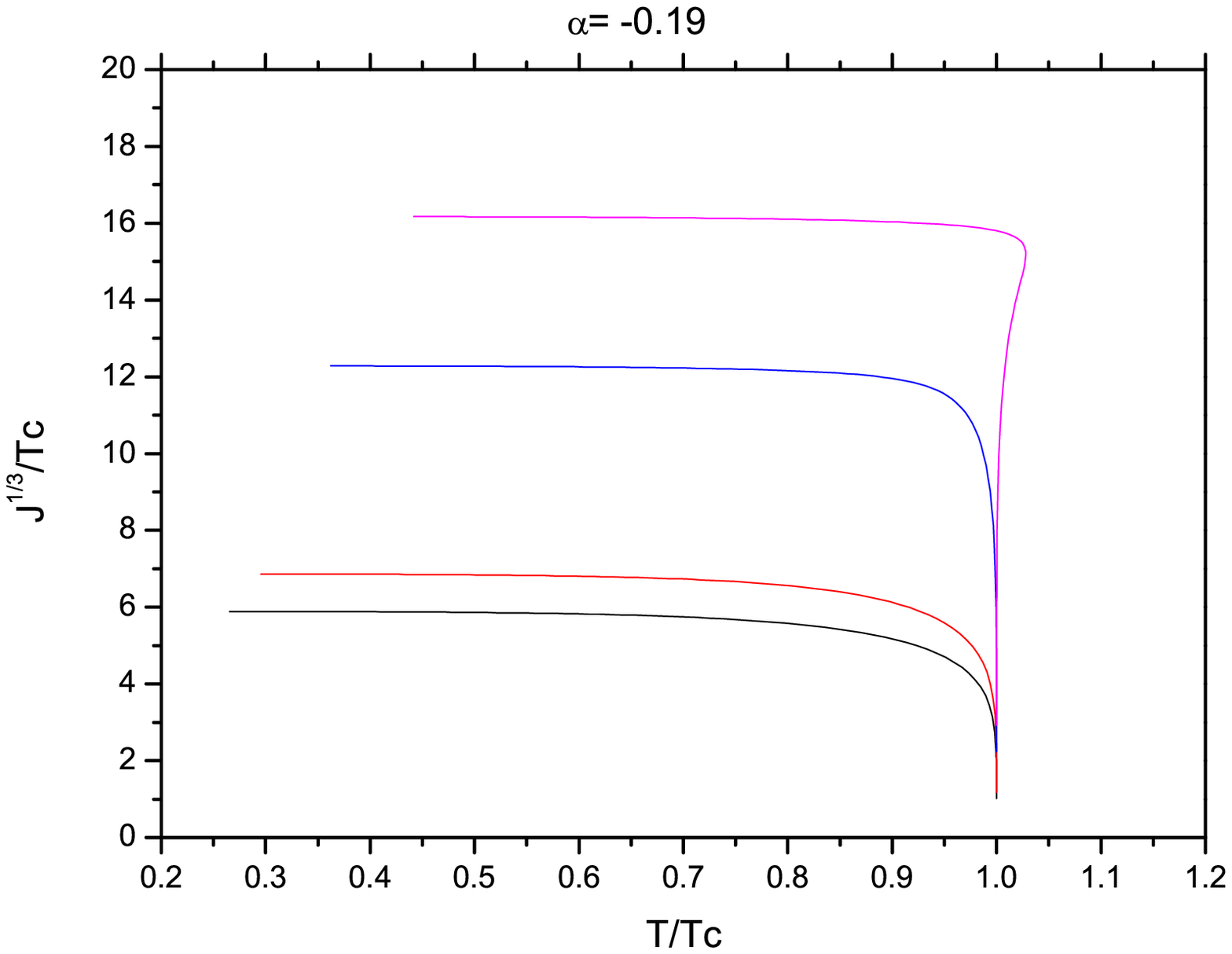}\hspace{-1.5cm}
\includegraphics[width=8.8cm] {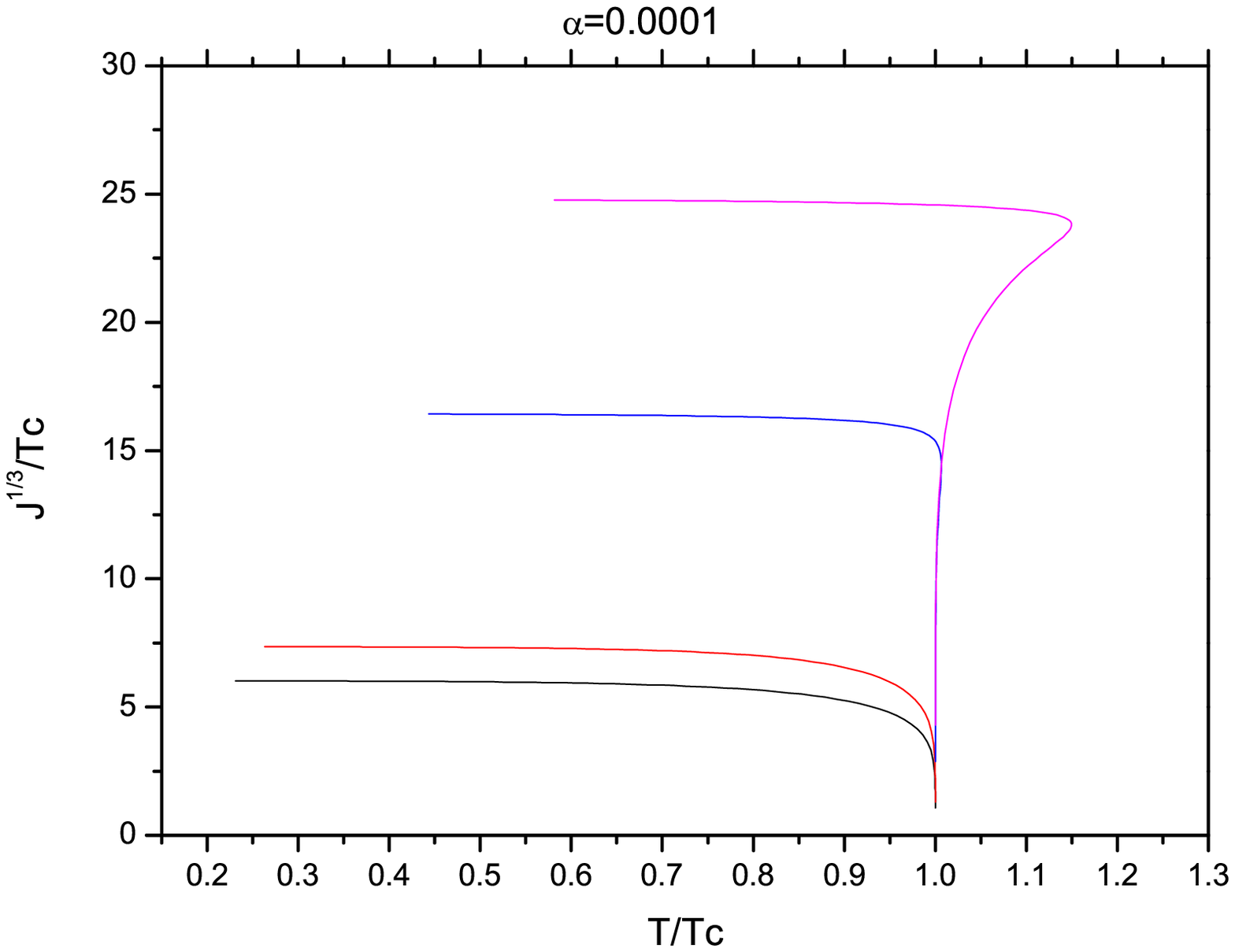}\\
\includegraphics[width=8.8cm] {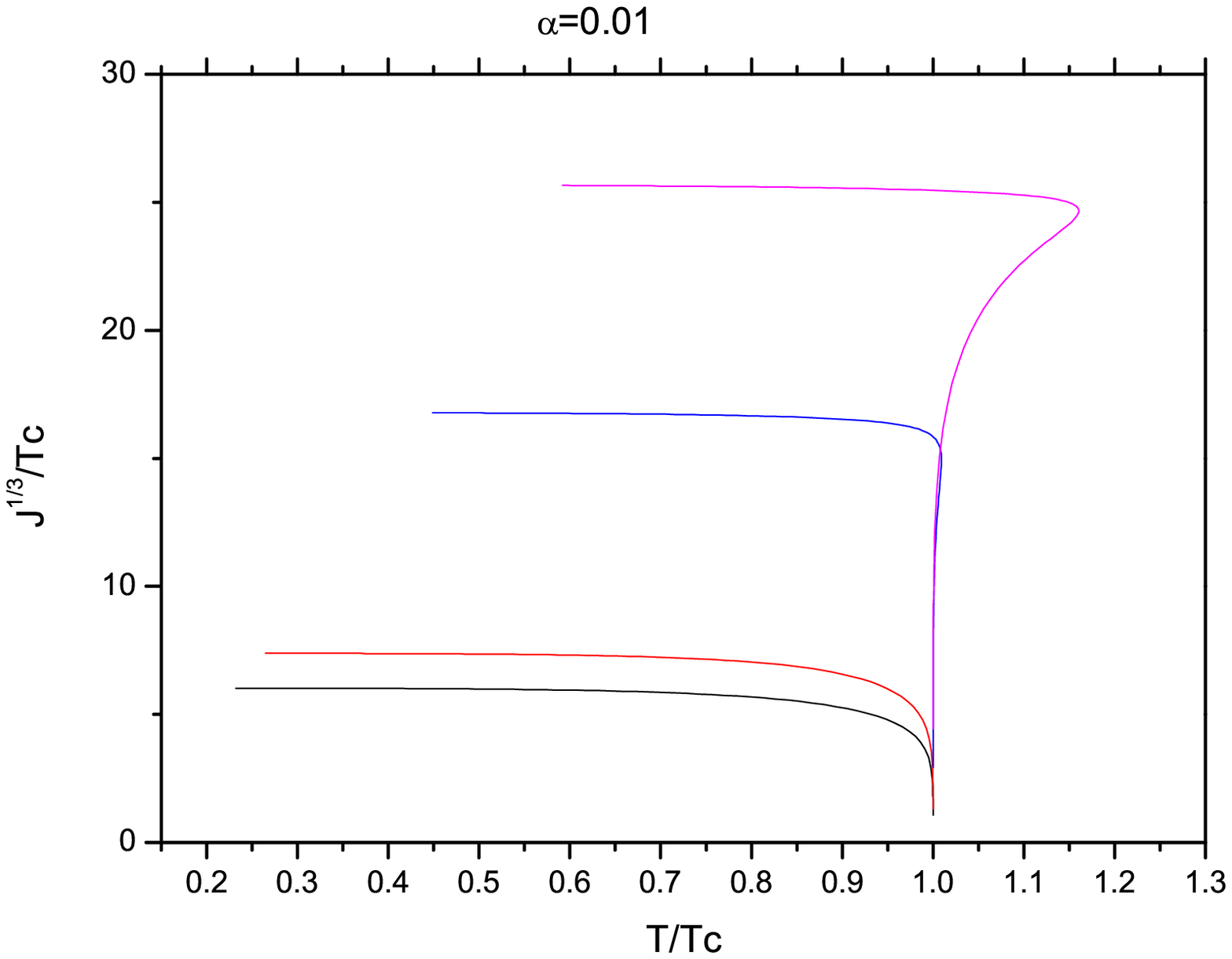}\hspace{-1.5cm}
\includegraphics[width=8.8cm] {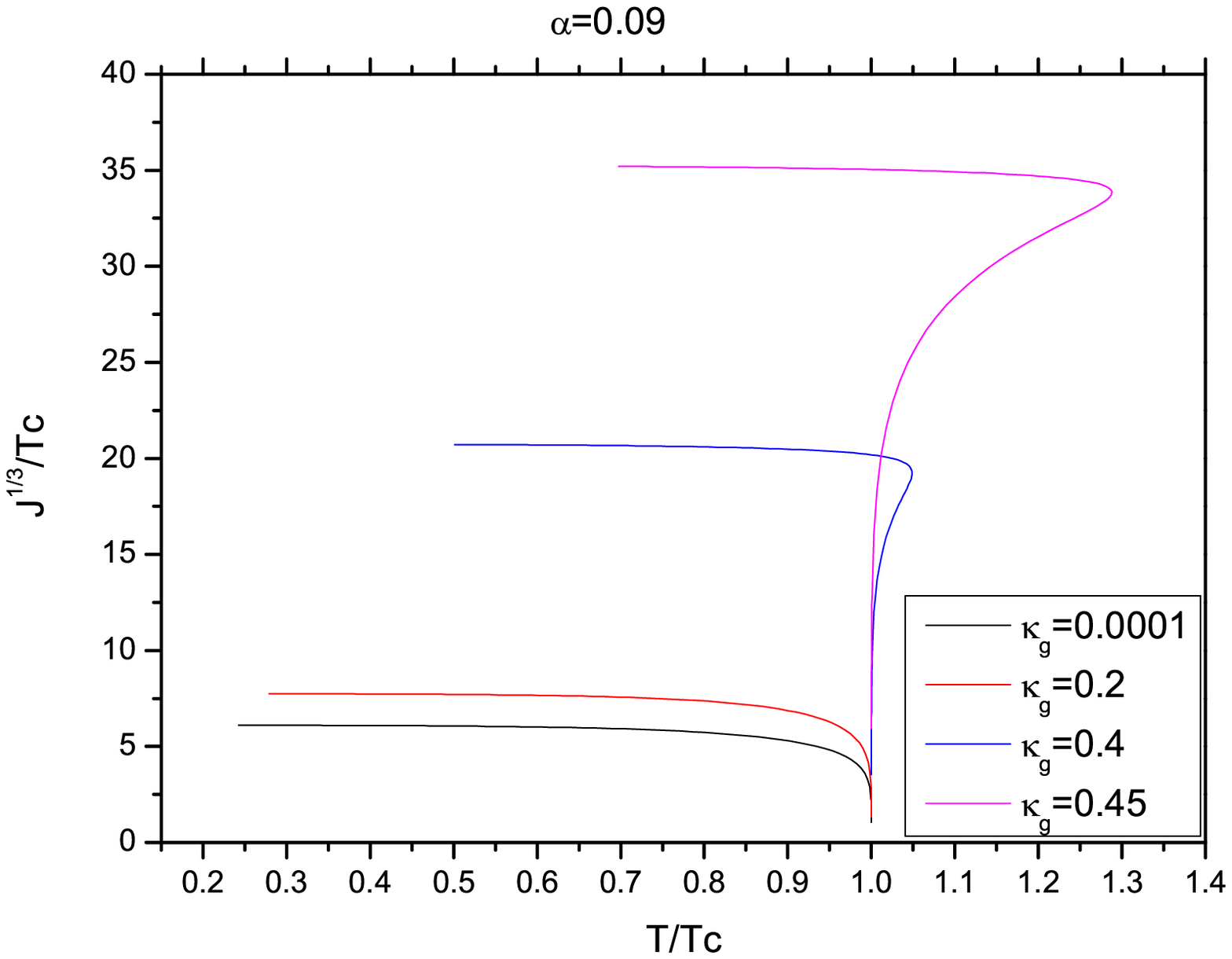}
\caption{\label{Condensation} The condensation values of vector
operator $J$ versus temperature for different $\al$ and different
$\ka_g$. The black, red, blue and pink curves correspond to
$\ka_g=0.0001, 0.2, 0.4$ and $0.45$, respectively.}
\end{figure}

\begin{table}
\caption{\label{tablereim} The condensation values of $J$ for
different $\al$ and $\ka_g$, numbers in boldface represent first
order phase transition}
\begin{center}
\begin{tabular}{|c||c|c|c|c|}
  % after \\: \hline or \cline{col1-col2} \cline{col3-col4} ...
  \hline
   \backslashbox{$\ka_g$}{$\al$} & $-0.19$  & 0.0001 & 0.01  & 0.09 \\
   \hline\hline
  0.0001 & 5.89103  & 6.01942 & 6.02764& 6.10352 \\
  \hline
  0.2 & 6.86288  & 7.35303 & 7.38871 & 7.74425 \\
  \hline
  0.4 & 12.28375  & \bf{16.42425} &\bf{16.77913} & \bf{20.71994} \\
  \hline
  0.45 & \bf{16.17033}  & \bf{24.76784} &\bf{25.65623}& \bf{35.20014} \\
  \hline
\end{tabular}
\end{center}
\end{table}

Fig.~\ref{Condensation} shows the condensation value of vector
operator $J$ for different Gauss-Bonnet coefficients and different
matter field couplings. We can see that the condensation value grows
if the Gauss-Bonnet coefficient grows or the matter field coupling
grows. Table.~\ref{tablereim} lists all the explicit condensation
data of Fig.~\ref{Condensation} .

Take the bottom-left plot in Fig.~\ref{Condensation} as an example
({\it i.e.}, the plot with $\al=0.01$), when we decrease the
temperature, the condensation values for black ($\ka_g=0.0001$) and
red ($\ka_g=0.2$) curve will emerge from zero at some critical
temperature $T_c$. When we keep on cooling down the system, those
condensation values will continuously tend to some constant values.
Besides, when $T\sim T_c$ the condensation will take a mean-field
theory critical exponent $1/2$ in the form of $J\propto
(1-T/T_c)^{1/2}$. These are the second order phase transitions which
have been explored in our previous paper \cite{Cai:2010cv}. However,
for the blue ($\ka_g=0.4$) and pink ($\ka_g=0.45$) lines the
behavior is much different from the above two curves. We see that
these curves will bend to the right for $T>T_c$ and there will be
two condensation values for $T>T_c$. However, when $T\ll T_c$ they
will also tend to some constant values. The mean-field theory
critical exponent $1/2$ never exists for these condensations. We
will argue in Sec.~\ref{sect:thermodynamics} that these peculiar
condensation behaviors are features of first order phase transitions
and the real critical temperature for these phase transitions in not
exactly $T_c$.

\begin{figure}
\includegraphics[width=10cm] {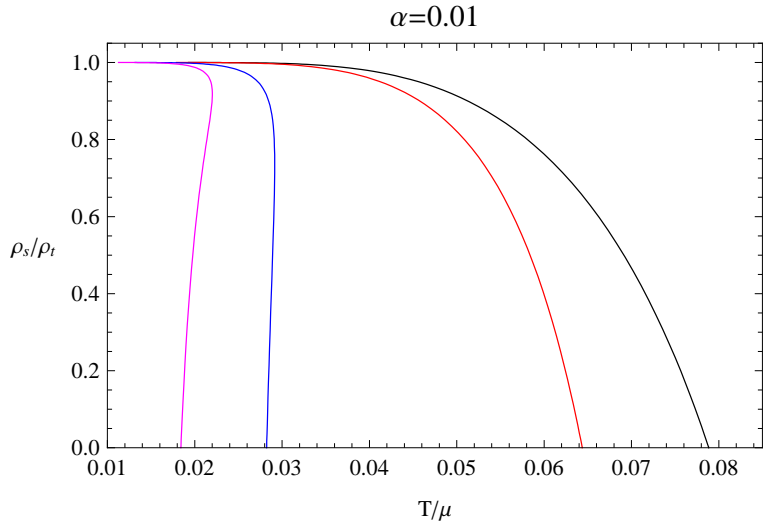}
\caption{\label{rhoratio} The ratio of superconducting charge
density and the total charge density versus the dimensionless
temperature $T/\mu$ at $\al=0.01$. The black, red, blue and pink
curves  correspond to $\ka_g=0.0001, 0.2, 0.4$ and $0.45$,
respectively.}
\end{figure}
In the holographic p-wave superconductors, the normal charge density
$\rho_n$ is $\phi_H^{(1)}$ near the horizon, while the total charge
density is $\rho_t=2\phi_B^{(2)}=2\rho$ where the factor $2$ appears
due to the scaling behavior of $\phi$ on the infinite boundary. The
superconducting charge density is defined as $\rho_s=\rho_t-\rho_n$
\cite{Gubser:2008wv,Cai:2010cv}. We plot the ratio $\rho_s/\rho_t$
in Fig.~\ref{rhoratio}. For the black and red curves which are of
second order phase transitions, the intersecting points where the
curves meet the horizontal axis represent the critical temperatures
$T_c/\mu$ at which the superconducting phase occurs. However, for
the blue and pink curves which are of first order phase transitions
(see Sec.~\ref{sect:thermodynamics}), the intersecting points are
not the critical temperature for the phase transitions. But the real
critical temperature $T_{c}^R$ for the first order phase transition
is  a little above the intersecting points $T_c$. From
Sec.~\ref{sect:thermodynamics}, we can find the values of
$T_c^R/\mu$ for $\al=0.01$ and varying $\ka_g$s, which are listed in
Table.~{\ref{tcmu}}. Despite of the first and second order phase
transitions, we still see that the dimensionless critical
temperature $T_c^R/\mu$ decreases when $\ka_g$ grows, which reflects
that the stronger the matter field couples to the gravity the harder
the phase transition occurs. For sufficiently low temperature, all
the curves will tend to one, which reflects the fact that the
superconducting charge is dominated in the total charge.

\begin{table}
\caption{\label{tcmu} The quantity of $T_c^R/\mu$ for different
$\ka_g$ while fixing $\al=0.01$. Boldfaces represent first order
phase transition.}
\begin{tabular}{|c|c|}
  \hline
  % after \\: \hline or \cline{col1-col2} \cline{col3-col4} ...
  $\ka_g$ & $T_c^R/\mu$ \\
  \hline\hline
  0.0001 & 0.07885 \\
  \hline
  0.2 & 0.06437 \\
  \hline
  0.4 & {\bf0.02880} \\
  \hline
  0.45 & {\bf0.02063} \\
  \hline
\end{tabular}
\end{table}

\section{Thermodynamics}
\label{sect:thermodynamics}

\subsection{Euclidean action and counter-term methods}
From the AdS/CFT correspondence, a non-extremal black hole
corresponds to a thermal equilibrium state at the boundary. The
Hawking temperature of black hole is the temperature of the boundary
field. In the following we will work in the grand canonical ensemble
with the fixed value of chemical potential $\mu$. The partition
function of the bulk theory is
 \be Z=e ^{-I_E[g_*]},\ee
 where $I_E[g_*]$ is the Euclidean action evaluated on the on-shell
 value of $g$. Because of the Euclidean action, the compactified time direction
 has a period $1/T$. The free energy now is
 \be \Om=-T\log Z=T\ I_E[g_*].\ee
  In the path integral, the Euclidean on-shell action
 should additionally include the Gibbons-Hawking surface term to
 give the correct Dirichlet variational problem and some other
 boundary counter-terms to render the action finite \cite{Balasubramanian:1999re}.
 In the computation we introduce a hypersurface at large but
 finite $r=r_{bdy}$ as the boundary, and then calculate the on-shell action by
 putting $r_{bdy}\rightarrow \infty$.

The bulk action evaluated on the on-shell values is
\be
I_{\text{on-shell}}^{\text{bulk}}&=&\frac{V}{T\
\ka_5^2}\bigg[\frac{r^2 N \sigma  \left(r
   f\right)'}{ f}\\\nno&&-\frac{\alpha
   N \left(rf\right)' }{f^3}\bigg(2f^2 ((r\si N)'+r\si'N)-r f' \left(r
   \sigma  N'+2 N \left(r
   \sigma \right)'\right) f-4 r^2 N \sigma
    f'^2\bigg)\bigg]\bigg{|}_{r= r_{bdy}},
\ee
 where $V/T=\int dt d^3x$ is the volume of the (3+1)-dimensional
 hypersurface. The usual Gibbons-Hawking surface term is \be
I^{(1)}_{GH}=-\frac{1}{\ka_5^2}\int d^4x \sqrt{-\ga}
K=-\frac{V}{T\ka_5^2}(3r^2N\si+\frac{r^3\si
N'}{2}+r^3N\si')\bigg|_{r=r_{bdy}},\ee where $\ga$ is the induced
metric on the $r=r_{bdy}$ hypersurface,
$n^{\mu}=\sqrt{N(r)}\dl^{\mu,r}$ is the outward-pointing normal
vector to the hypersurface and $K=K^{\mu}_{\mu}$ is the trace of the
extrinsic curvature $K_{\mu\nu}=\nabla_{(\mu}n_{\nu)}$.

For the GB term there is also a generalized Gibbons-Hawking term
\cite{Myers:1987yn,Davis:2002gn}
 \be I^{(2)}_{GH}&=&-\frac{\al}{\ka_5^2}\int
 d^4x\sqrt{-\ga}(J-2G_{ij}K^{ij})\\
 &=&\frac{\al V N}{T \ka_5^2 f^3}
 \bigg(\left(3 r \sigma
   N'+2 N \left(\sigma +3 r \sigma
   '\right)\right) f^3-\nno\\
   &&
   3 r^2 f'^2
   \left(r \sigma  N'+2 N \left(\sigma
   +r \sigma '\right)\right) f-4 r^3
   N \sigma  f'^3\bigg)
   \bigg|_{r= r_{bdy}},\nno\ee
where $G_{ij}$ is the Einstein tensor of the metric $\ga_{ij}$ and
$J=\ga^{ij}J_{ij}$ with
 \be
 J_{ij}=\frac{1}{3}(2KK_{ik}K^{k}_{j}+K_{kl}K^{kl}K_{ij}-2K_{ik}K^{kl}K_{lj}-K^2K_{ij}).\ee

No additional counter-terms for matter fields are necessary because
the fields fall off sufficiently near the boundary
\cite{Hartnoll:2009sz}.\footnote{If one works in the canonical
ensemble, the charge density $\rho$ should be fixed and we should
add an additional boundary term to the Euclidean action as $\Dl
I_E\propto \int d^4x \sqrt{-\ga} \text{tr}(n^{\mu}F^a_{\mu\nu}A^{a
\nu})$. \cite{Gibbons:1976ue}} Besides, in our metric ansatz the
scalar curvature $\mathrm{R}$ for the hypersurface is zero. So the
simplest counter-term is
\cite{Kofinas:2006hr,Brihaye:2008kh,Astefanesei:2008wz,Brihaye:2008xu}
 \be I_{ct}=\frac{1}{\ka_5^2}\int
 d^4x\sqrt{-\ga}
 \frac{3}{L_c}(\frac{2+U}{3})=\frac{V}{T\ka_5^2}\sqrt{N}\si
 r^3\frac{2+U}{L_c}\bigg|_{r=r_{bdy}}.\ee
Thus the total on-shell Euclidean action $I_E[g_*]$ is
 \be I_{E}[g_*]=I_{\text{on-shell}}^{\text{
 bulk}}+I^{(1)}_{GH}+I^{(2)}_{GH}+I_{ct}.\ee
Then the free energy is
 \be \Om=T\ I_E[g_*]=T\ (I_{\text{on-shell}}^{\text{
 bulk}}+I^{(1)}_{GH}+I^{(2)}_{GH}+I_{ct}).\ee

 \subsection{Phase diagrams of free energy and entropy}

\begin{figure}
\includegraphics[width=8.cm] {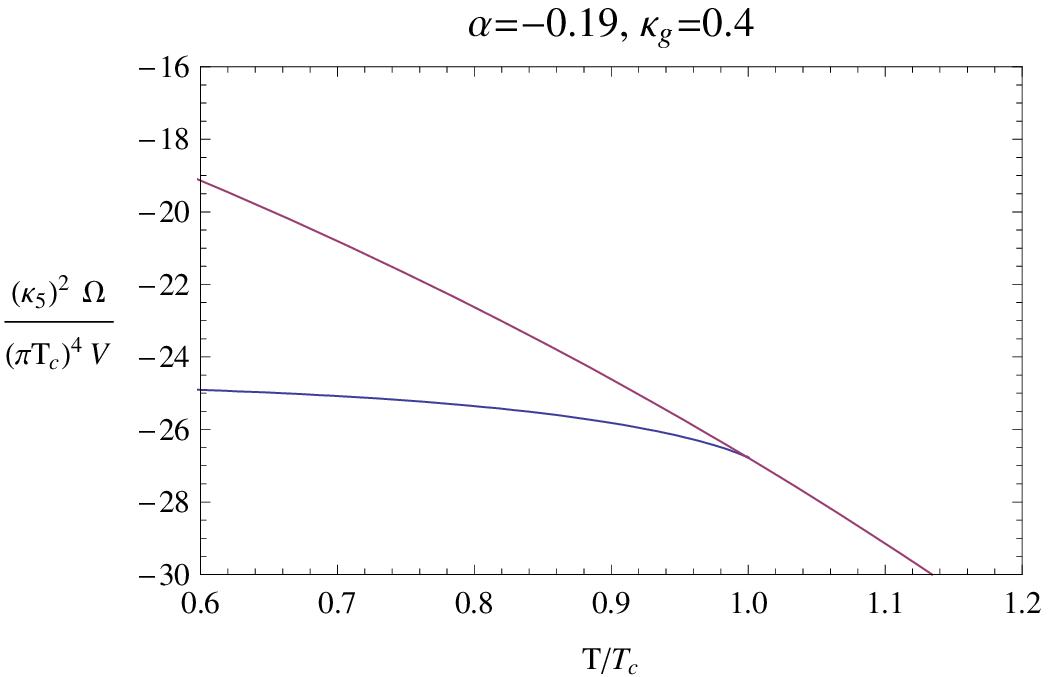}
\includegraphics[width=8.cm] {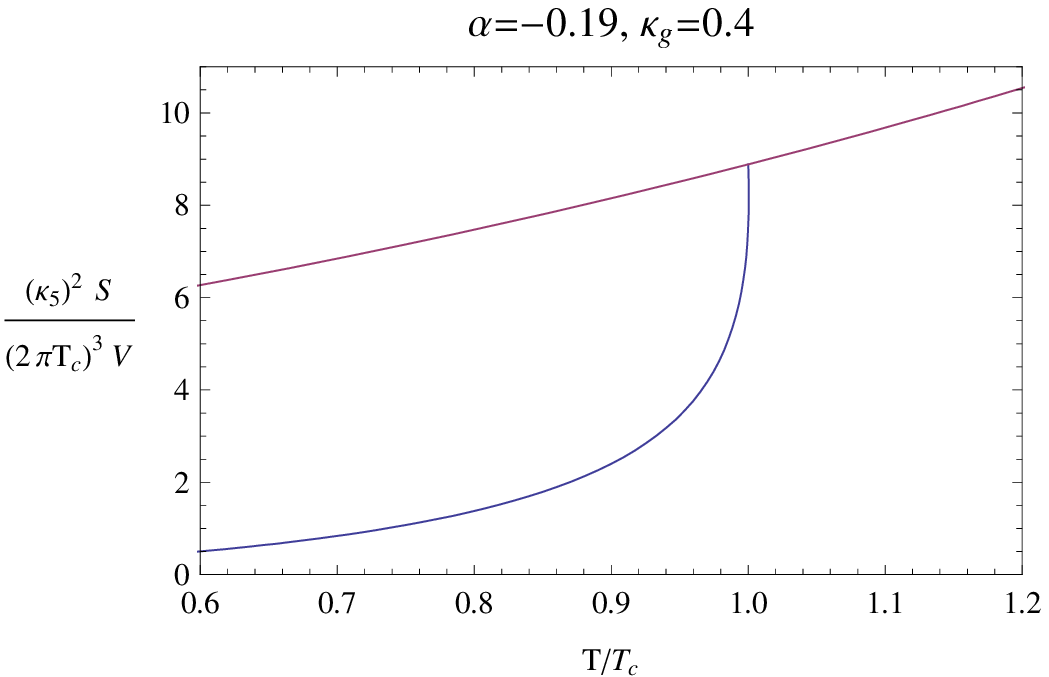}\\
\includegraphics[width=8.cm] {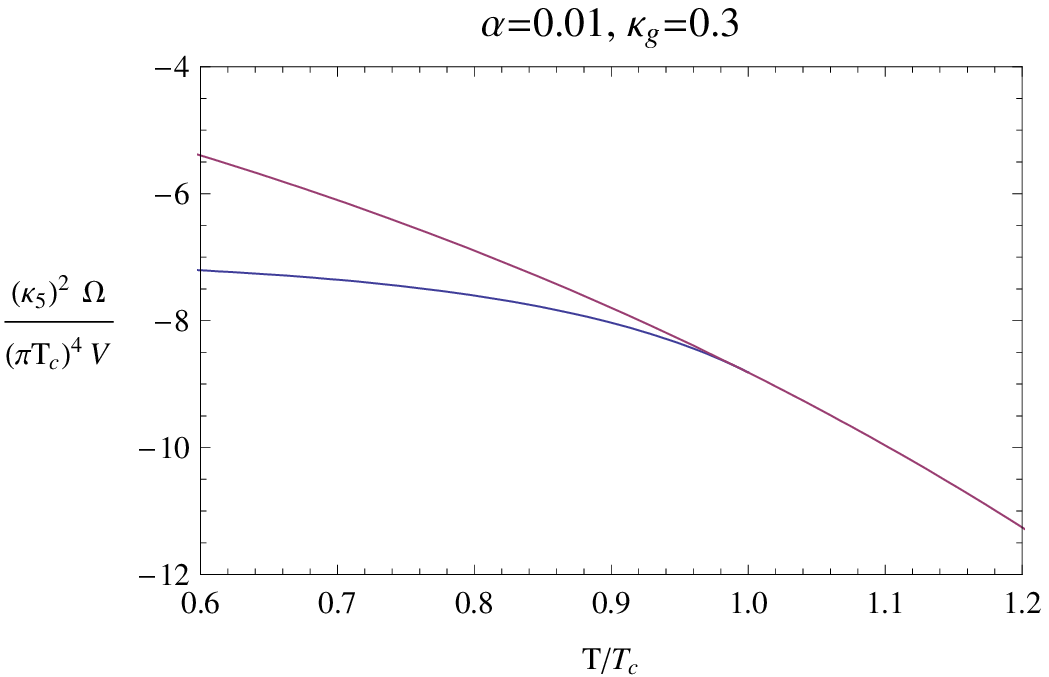}
\includegraphics[width=8.cm] {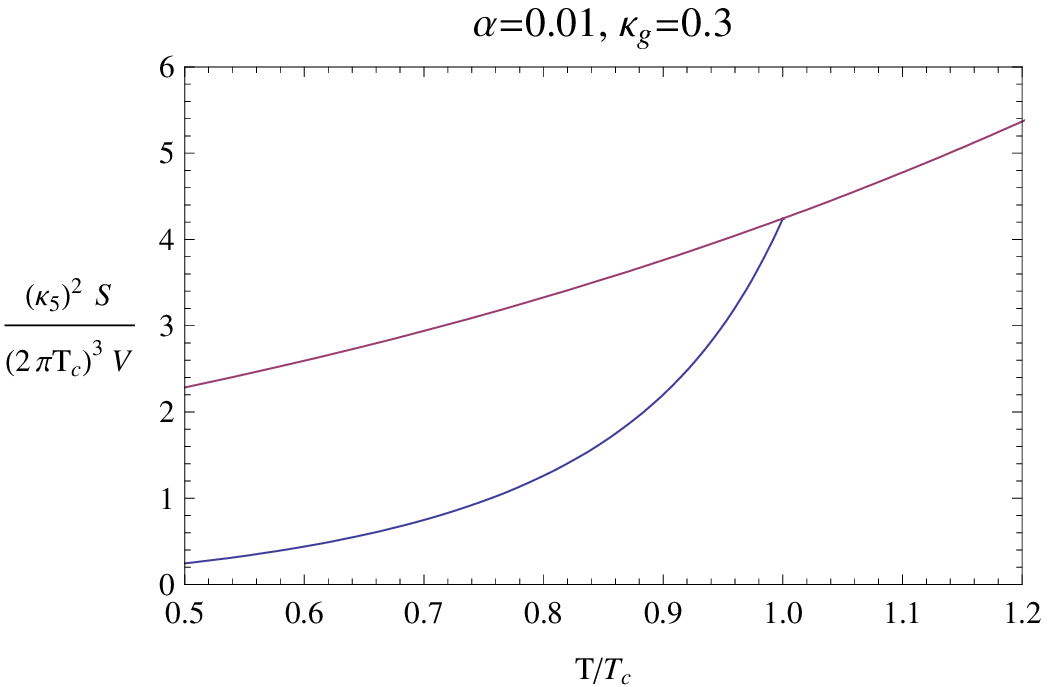}
\\
\includegraphics[width=8.cm] {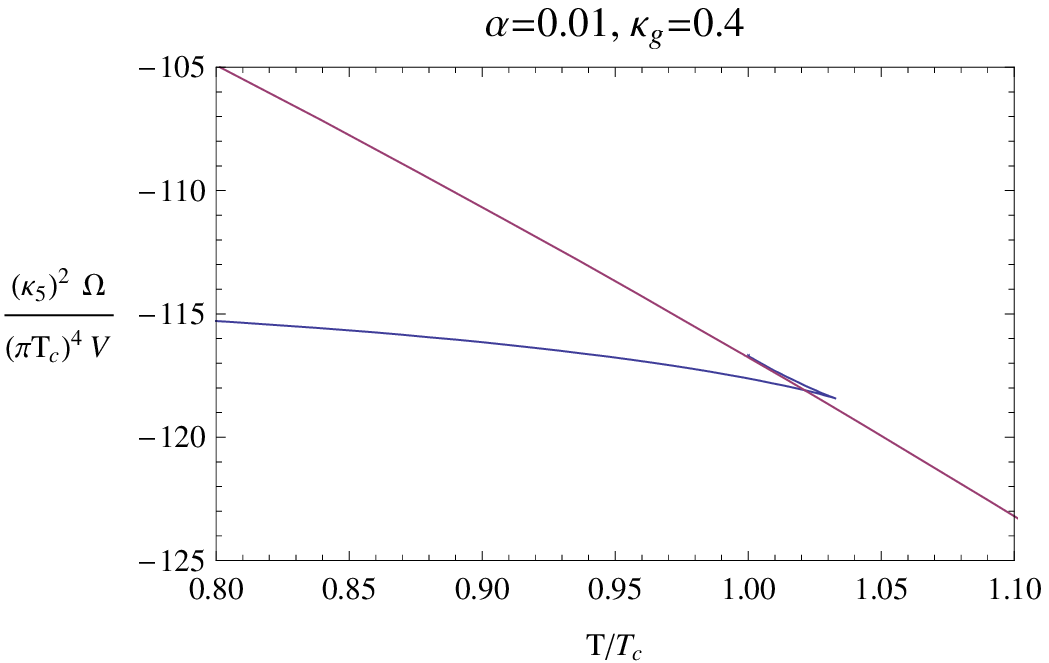}
\includegraphics[width=8.cm] {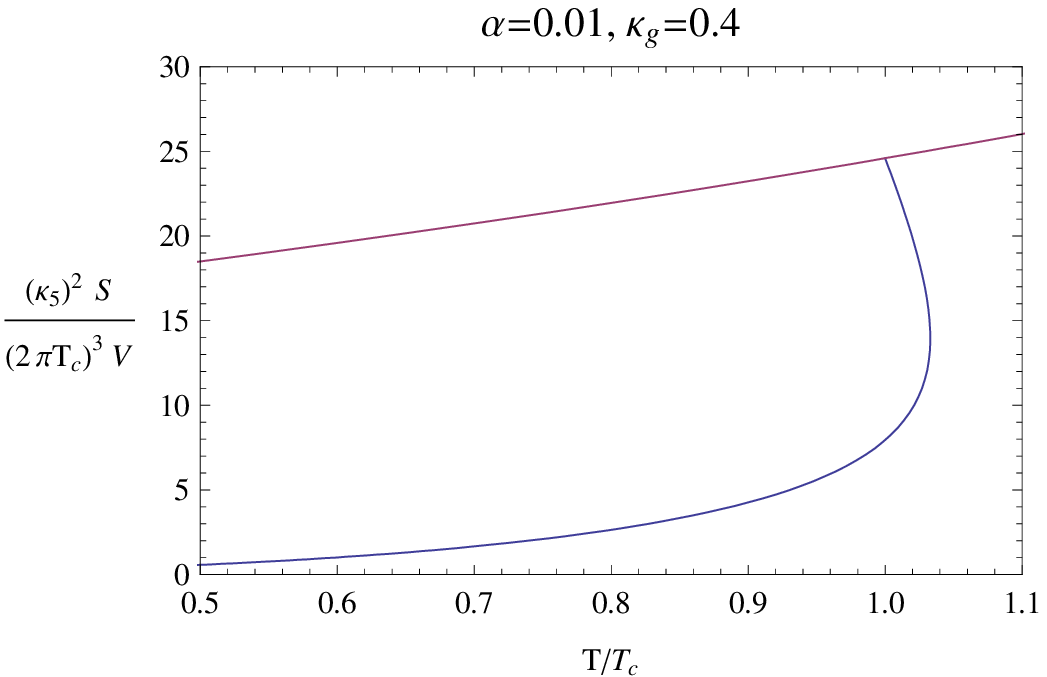}
\\
\includegraphics[width=8.cm] {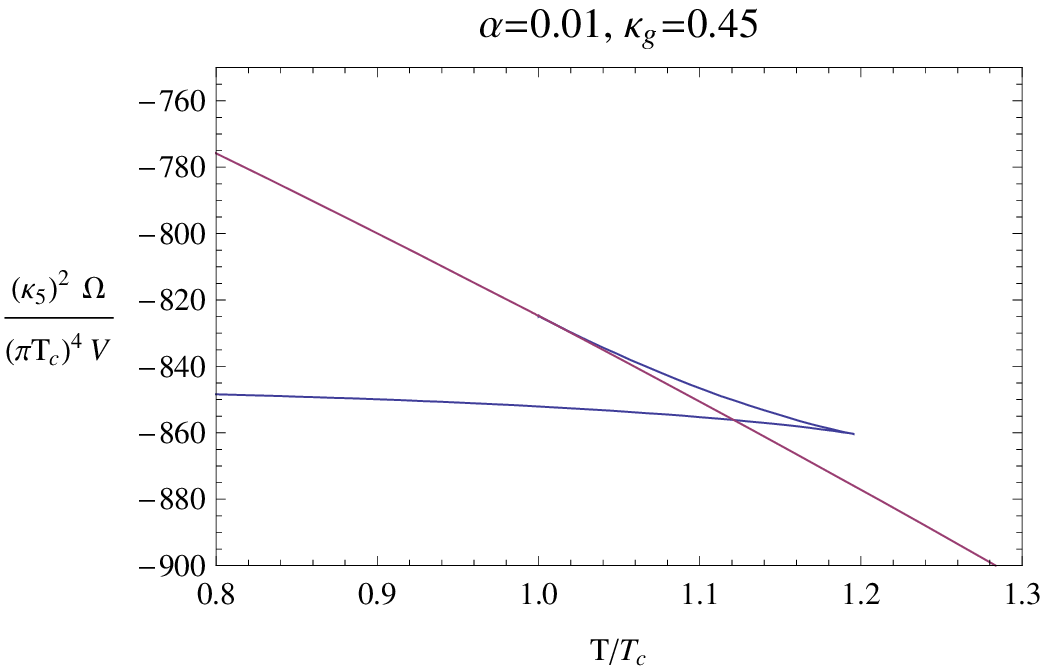}
\includegraphics[width=8.cm] {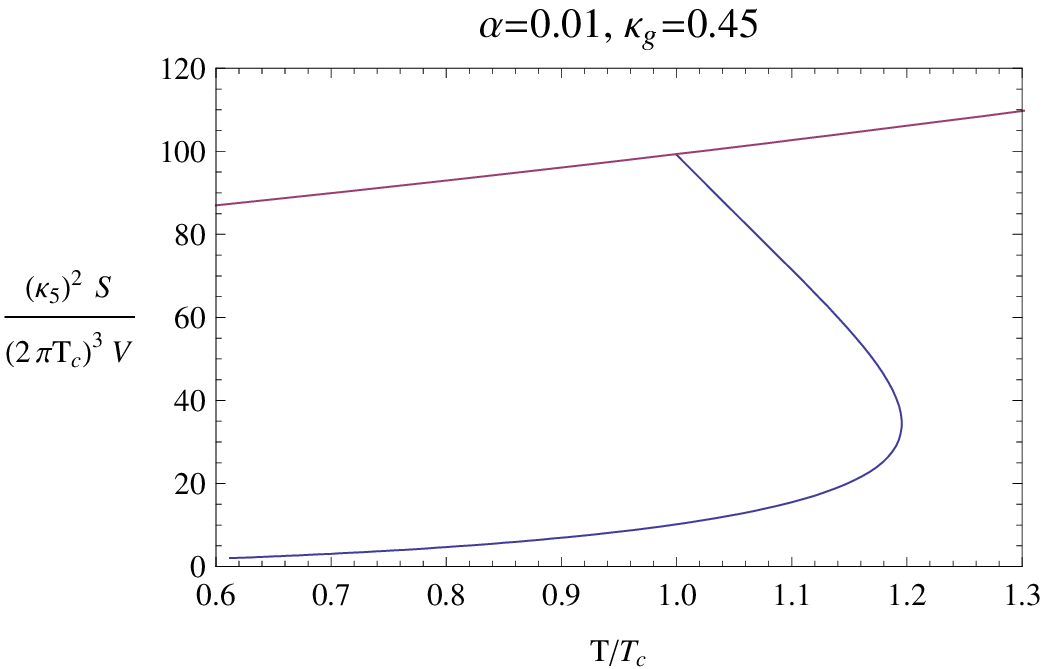}
\caption{\label{thermal} The free energy and entropy versus
temperature for different $\al$ and $\ka_g$. The purple line is for
the charged GB-AdS black hole ({\it i.e.}, $w(r)=0$) while the blue
line is for the superconducting solutions ({\it i.e.}, $w(r)\neq0$)
. }
\end{figure}

In this subsection, we will discuss the numerical results of the
free energy $\Om$ and the entropy $S$. In Fig.~\ref{thermal}, we
plot the free energy and entropy for some typical values of $\al$
and $\ka_g$.

From the first two up-left plots in Fig.~\ref{thermal}, we find that
for $\al=-0.19, \ka_g=0.4$ and $\al=0.01, \ka_g=0.3$, the charged
GB-AdS solution (purple line) (\ref{rnads}) exists for all the
temperature $T>T_c$ and $T\leq T_c$. On the contrary, the
superconducting solution (blue line) exists only for $T\leq T_c$.
But for $T<T_c$ the free energy of superconducting solution is
smaller than that of the charged GB-AdS solution, which means that
the superconducting solution is thermodynamically favored, compared
to the charged GB-AdS solution. This represents that when $T$
decreases across $T_c$, a phase transition occurs when the bulk
solution going from the charged GB-AdS solution to a solution with
nonzero vector hair which induces an order parameter of the p-wave
holographic superconductor. In addition, we see that at $T=T_c$ the
purple line and blue line are not only continuous but also $C^1$
differentiable which can be seen from the plots of entropy $S$
because $S=-\pa \Om/\pa T$, see the first two up-right plots in
Fig.~\ref{thermal}. The entropy for these two solutions at $T=T_c$
is continuous but not differentiable. So according to Ehrenfest's
classification of phase transitions, these phase transitions are
second order for $\al=-0.19, \ka_g=0.4$ and $\al=0.01,
 \ka_g=0.3$. This is consistent with our remarks in the previous
 section.

 From the two down-left plots of Fig.~\ref{thermal}, {\it i.e.}, plots with
 $\al=0.01, \ka_g=0.4$
 and $\al=0.01, \ka_g=0.45$, the behavior
 of free energy is dramatically different from the one
 we mentioned above, there is a characteristic ``swallowtail" shape
 of the free energy indicating a first order phase transition. Consider the
  $\al=0.01, \ka_g=0.45$
 case for example. When we decrease the temperature, entering the
 figure along the purple line from the right, we reach the
 temperature $T\approx1.2T_c$ where new solutions appear (the blue
 curves
 represent the new solutions). However, for now the charged GB-AdS solution is thermodynamically
 favored because it has a lower free energy. If we keep on cooling
 down the system we will still remain in the charged GB-AdS solution until
 $T\approx1.12T_c$. Then for $T<1.12T_c$, the superconducting solution
 will have a lower free energy than the charged GB-AdS solution. Therefore, superconducting
 solution will be  thermodynamically favored, compared to the charged GB-AdS
 solution in the range $T<1.12T_c$. So the exact critical
 temperature for this kind of phase transition is $T_c^R=1.12T_c$. This is
 the reason why the quantities for $\ka_g=0.4$ and $\ka_g=0.45$ in Table.~\ref{tcmu}
 are not the exact values which the intersecting points denote.
 This is a first order phase transition due to the non-differentiable free energy
 at $T=1.12T_c$. From the point view of entropy, we
 see that the entropy will jump from the purple curve to the lowest part of the
  blue curve at $T=1.12T_c$ (see the down-right plot in Fig.~\ref{thermal}.
  The entropy is not continuous which
 also reveals the feature of the first order phase transition. Therefore, from the
 plots in Fig.~\ref{thermal},
 we can read the real critical temperature for the first order phase transition
 as $T=1.02T_c$ when $\al=0.01,\ka_g=0.4$ and $T=1.12T_c$ when $\al=0.01, \ka_g=0.45$,
 respectively.

\section{\bf Conclusions}
\label{sect:conclusion}

\begin{figure}
\includegraphics[width=10cm] {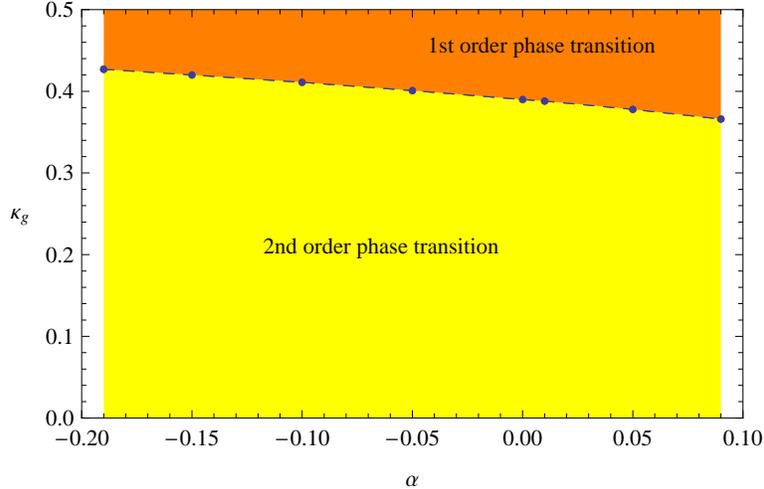}
\caption{\label{kgc} Classification of phase transition versus
 $\al\ (-0.19\leq\al\leq0.09)$ and
$\ka_g\ (\ka_g\geq0)$. The order of phase transition depends on the
coefficients $\al$ and $\ka_g$. The yellow region below the dashed
curve is of the second order phase transition while the orange
region is of the first order phase transition.}
\end{figure}

 In a previous paper~\cite{Cai:2010cv}, we studied holographic p-wave superconductors
 within Gauss-Bonnet gravity in the probe limit. In this paper, we  continued this
 study by including back-reaction of Yang-Mills field. We found that both the Gauss-Bonnet
 coefficient and back-reaction will make the superconducting condensation
  difficult. This difficulty can be seen both from the
 growing condensation values and the decreasing critical temperatures.
By studying the thermodynamics of the system in grand canonical
ensemble, we found two kinds of phase transitions of the holographic
p-wave superconductors. Note that in the probe limit, the
superconducting phase transition is always second order.  With
back-reaction, we found that when fixing the Gauss-Bonnet
coefficient, there was a critical value for the matter field
couplings $\ka_{g(c)}$(see Fig.~\ref{kgc}). If $\ka_g<\ka_{g(c)}$,
the phase transition is second order (yellow region), however, if
$\ka_g>\ka_{g(c)}$, the phase transition becomes first order (orange
region). It was found that the stronger back-reaction will not only
make the condensation value bigger but also will change the order of
the phase transition. This is consistent with the conclusions of
\cite{Ammon:2009xh}. On the contrary, if we fixed the matter field
couplings $\ka_g$, the Gauss-Bonnet couplings would change the order
of the phase transition just for a small range of $\ka_g$, {\it
i.e.}, $0.366\leq\ka_g\leq0.427$. However, out of this range, the
Gauss-Bonnet term would not change the order of the phase
transition. Therefore, we may conclude that although both the
Gauss-Bonnet coefficient and back-reaction will make the p-wave
condensation hard, the back-reaction plays a major role in changing
the order of the phase transition.

%=========================================================================
\acknowledgments
 ZYN and HQZ would like to thank Bin Hu, Ya-Peng Hu, Huai-Fan Li, Zhi-Yuan Xie and Yun-Long
 Zhang for helpful discussions.  This work was supported in part by a grant from
Chinese Academy of Sciences and  in part by the National Natural
Science Foundation of China under Grant Nos. 10821504, 10975168 and
11035008, and by the Ministry of Science and Technology of China
under Grant No. 2010CB833004.
%=========================================================================

%%%%%%%%%%%%%%%%%%%%%%%%%%%%%%%%%%%%%%%%%%%%%%%%%%%%%%%%%%%%%%%%%%%%%%%%%%%%
%                      REFERENCES                                          %
%%%%%%%%%%%%%%%%%%%%%%%%%%%%%%%%%%%%%%%%%%%%%%%%%%%%%%%%%%%%%%%%%%%%%%%%%%%%
%\newpage


\begin{thebibliography}{99}
\baselineskip 12pt

%\cite{Maldacena:1997re}
\bibitem{Maldacena:1997re}
  J.~M.~Maldacena,
  %``The large N limit of superconformal field theories and supergravity,''
  Adv.\ Theor.\ Math.\ Phys.\  {\bf 2}, 231 (1998)
  [Int.\ J.\ Theor.\ Phys.\  {\bf 38}, 1113 (1999)]
  [arXiv:hep-th/9711200].
  %%CITATION = IJTPB,38,1113;%%

%\cite{Gubser:1998bc}
\bibitem{Gubser:1998bc}
  S.~S.~Gubser, I.~R.~Klebanov and A.~M.~Polyakov,
  %``Gauge theory correlators from non-critical string theory,''
  Phys.\ Lett.\  B {\bf 428}, 105 (1998)
  [arXiv:hep-th/9802109].
  %%CITATION = PHLTA,B428,105;%%

%\cite{Witten:1998qj}
\bibitem{Witten:1998qj}
  E.~Witten,
  %``Anti-de Sitter space and holography,''
  Adv.\ Theor.\ Math.\ Phys.\  {\bf 2}, 253 (1998)
  [arXiv:hep-th/9802150].
  %%CITATION = 00203,2,253;%%

%\cite{Aharony:1999ti}
\bibitem{Aharony:1999ti}
  O.~Aharony, S.~S.~Gubser, J.~M.~Maldacena, H.~Ooguri and Y.~Oz,
  %``Large N field theories, string theory and gravity,''
  Phys.\ Rept.\  {\bf 323}, 183 (2000)
  [arXiv:hep-th/9905111].
  %%CITATION = PRPLC,323,183;%%

  %\cite{Policastro:2001yc}
\bibitem{Policastro:2001yc}
  G.~Policastro, D.~T.~Son and A.~O.~Starinets,
  %``The shear viscosity of strongly coupled N = 4 supersymmetric Yang-Mills
  %plasma,''
  Phys.\ Rev.\ Lett.\  {\bf 87}, 081601 (2001)
  [arXiv:hep-th/0104066].
  %%CITATION = PRLTA,87,081601;%%

%\cite{Kovtun:2003wp}
\bibitem{Kovtun:2003wp}
  P.~Kovtun, D.~T.~Son and A.~O.~Starinets,
  %``Holography and hydrodynamics: Diffusion on stretched horizons,''
  JHEP {\bf 0310}, 064 (2003)
  [arXiv:hep-th/0309213].
  %%CITATION = JHEPA,0310,064;%%

%\cite{Buchel:2003tz}
\bibitem{Buchel:2003tz}
  A.~Buchel and J.~T.~Liu,
  %``Universality of the shear viscosity in supergravity,''
  Phys.\ Rev.\ Lett.\  {\bf 93}, 090602 (2004)
  [arXiv:hep-th/0311175].
  %%CITATION = PRLTA,93,090602;%%

%\cite{Kovtun:2004de}
\bibitem{Kovtun:2004de}
  P.~Kovtun, D.~T.~Son and A.~O.~Starinets,
  %``Viscosity in strongly interacting quantum field theories from black hole
  %physics,''
  Phys.\ Rev.\ Lett.\  {\bf 94}, 111601 (2005)
  [arXiv:hep-th/0405231].
  %%CITATION = PRLTA,94,111601;%%

  %\cite{Gubser:2008px}
\bibitem{Gubser:2008px}
  S.~S.~Gubser,
  %``Breaking an Abelian gauge symmetry near a black hole horizon,''
  Phys.\ Rev.\  D {\bf 78}, 065034 (2008)
  [arXiv:0801.2977 [hep-th]].
  %%CITATION = PHRVA,D78,065034;%%
%\cite{Hartnoll:2008vx}
\bibitem{Hartnoll:2008vx}
  S.~A.~Hartnoll, C.~P.~Herzog and G.~T.~Horowitz,
  %``Building a Holographic Superconductor,''
  Phys.\ Rev.\ Lett.\  {\bf 101}, 031601 (2008)
  [arXiv:0803.3295 [hep-th]].
  %%CITATION = PRLTA,101,031601;%%

  %\cite{Lee:2008xf}
\bibitem{Lee:2008xf}
  S.~S.~Lee,
  %``A Non-Fermi Liquid from a Charged Black Hole: A Critical Fermi Ball,''
  Phys.\ Rev.\  D {\bf 79} (2009) 086006
  [arXiv:0809.3402 [hep-th]].
  %%CITATION = PHRVA,D79,086006;%%


%\cite{Liu:2009dm}
\bibitem{Liu:2009dm}
  H.~Liu, J.~McGreevy and D.~Vegh,
  %``Non-Fermi liquids from holography,''
  arXiv:0903.2477 [hep-th].
  %%CITATION = ARXIV:0903.2477;%%

%\cite{Cubrovic:2009ye}
\bibitem{Cubrovic:2009ye}
  M.~Cubrovic, J.~Zaanen and K.~Schalm,
  %``String Theory, Quantum Phase Transitions and the Emergent Fermi-Liquid,''
  Science {\bf 325} (2009) 439
  [arXiv:0904.1993 [hep-th]].
  %%CITATION = SCIEA,325,439;%%

  %\cite{Gubser:2008wv}
\bibitem{Gubser:2008wv}
  S.~S.~Gubser and S.~S.~Pufu,
  %``The gravity dual of a p-wave superconductor,''
  JHEP {\bf 0811} (2008) 033
  [arXiv:0805.2960 [hep-th]].
  %%CITATION = JHEPA,0811,033;%%

  %\cite{Basu:2009vv}
\bibitem{Basu:2009vv}
  P.~Basu, J.~He, A.~Mukherjee and H.~H.~Shieh,
  %``Hard-gapped Holographic Superconductors,''
  Phys.\ Lett.\  B {\bf 689} (2010) 45
  [arXiv:0911.4999 [hep-th]].
  %%CITATION = PHLTA,B689,45;%%


%\cite{Ammon:2009xh}
\bibitem{Ammon:2009xh}
  M.~Ammon, J.~Erdmenger, V.~Grass, P.~Kerner and A.~O'Bannon,
  %``On Holographic p-wave Superfluids with Back-reaction,''
  Phys.\ Lett.\  B {\bf 686} (2010) 192
  [arXiv:0912.3515 [hep-th]].
  %%CITATION = PHLTA,B686,192;%%

  %\cite{Manvelyan:2008sv}
\bibitem{Manvelyan:2008sv}
  R.~Manvelyan, E.~Radu and D.~H.~Tchrakian,
  %``New AdS non Abelian black holes with superconducting horizons,''
  Phys.\ Lett.\  B {\bf 677} (2009) 79
  [arXiv:0812.3531 [hep-th]].
  %%CITATION = PHLTA,B677,79;%%


%\cite{Weinberg:1986cq}
\bibitem{Weinberg:1986cq}
  S.~Weinberg,
  %``Superconductivity For Particular Theorists,''
  Prog.\ Theor.\ Phys.\ Suppl.\  {\bf 86} (1986) 43.
  %%CITATION = PTPSA,86,43;%%

  %\cite{Cai:2010cv}
\bibitem{Cai:2010cv}
  R.~G.~Cai, Z.~Y.~Nie and H.~Q.~Zhang,
  %``Holographic p-wave superconductors from Gauss-Bonnet gravity,''
  Phys.\ Rev.\  D {\bf 82} (2010) 066007
  [arXiv:1007.3321 [hep-th]].
  %%CITATION = PHRVA,D82,066007;%%


%\cite{Barclay:2010up}
\bibitem{Barclay:2010up}
  L.~Barclay, R.~Gregory, S.~Kanno and P.~Sutcliffe,
  %``Gauss-Bonnet Holographic Superconductors,''
  JHEP {\bf 1012} (2010) 029
  [arXiv:1009.1991 [hep-th]].
  %%CITATION = JHEPA,1012,029;%%

  %\cite{Gregory:2010yr}
\bibitem{Gregory:2010yr}
  R.~Gregory,
  %``Holographic Superconductivity with Gauss-Bonnet gravity,''
  arXiv:1012.1558 [hep-th].
  %%CITATION = ARXIV:1012.1558;%%

%\cite{Barclay:2010nm}
\bibitem{Barclay:2010nm}
  L.~Barclay,
  %``The Rich Structure of Gauss-Bonnet Holographic Superconductors,''
  arXiv:1012.3074 [hep-th].
  %%CITATION = ARXIV:1012.3074;%%


%\cite{Gregory:2009fj}
\bibitem{Gregory:2009fj}
  R.~Gregory, S.~Kanno and J.~Soda,
  %``Holographic Superconductors with Higher Curvature Corrections,''
  JHEP {\bf 0910} (2009) 010
  [arXiv:0907.3203 [hep-th]].
  %%CITATION = JHEPA,0910,010;%%

  %\cite{Pan:2009xa}
\bibitem{Pan:2009xa}
  Q.~Pan, B.~Wang, E.~Papantonopoulos, J.~Oliveira and A.~B.~Pavan,
  %``Holographic Superconductors with various condensates in
  %Einstein-Gauss-Bonnet gravity,''
  Phys.\ Rev.\  D {\bf 81} (2010) 106007
  [arXiv:0912.2475 [hep-th]].
  %%CITATION = PHRVA,D81,106007;%%

    %\cite{Brihaye:2010mr}
\bibitem{Brihaye:2010mr}
  Y.~Brihaye and B.~Hartmann,
  %``Holographic Superconductors in 3+1 dimensions away from the probe limit,''
  Phys.\ Rev.\  D {\bf 81} (2010) 126008
  [arXiv:1003.5130 [hep-th]].
  %%CITATION = PHRVA,D81,126008;%%


%\cite{Pan:2010at}
\bibitem{Pan:2010at}
  Q.~Pan and B.~Wang,
  %``General holographic superconductor models with Gauss-Bonnet corrections,''
  Phys.\ Lett.\  B {\bf 693}, 159 (2010)
  [arXiv:1005.4743 [hep-th]].
  %%CITATION = PHLTA,B693,159;%%

%\cite{Siani:2010uw}
\bibitem{Siani:2010uw}
  M.~Siani,
  %``Holographic Superconductors and Higher Curvature Corrections,''
  JHEP {\bf 1012} (2010) 035
  [arXiv:1010.0700 [hep-th]].
  %%CITATION = JHEPA,1012,035;%%

%\cite{Jing:2010cx}
\bibitem{Jing:2010cx}
  J.~Jing, L.~Wang, Q.~Pan and S.~Chen,
  %``Holographic Superconductors in Gauss-Bonnet gravity with Born-Infeld
  %electrodynamics,''
  arXiv:1012.0644 [gr-qc].
  %%CITATION = ARXIV:1012.0644;%%

  %\cite{Brigante:2007nu}
\bibitem{Brigante:2007nu}
  M.~Brigante, H.~Liu, R.~C.~Myers, S.~Shenker and S.~Yaida,
  %``Viscosity Bound Violation in Higher Derivative Gravity,''
  Phys.\ Rev.\  D {\bf 77}, 126006 (2008)
  [arXiv:0712.0805 [hep-th]];
  %%CITATION = PHRVA,D77,126006;%%
%\cite{Brigante:2008gz}
\bibitem{Brigante:2008gz}
  M.~Brigante, H.~Liu, R.~C.~Myers, S.~Shenker and S.~Yaida,
  %``The Viscosity Bound and Causality Violation,''
  Phys.\ Rev.\ Lett.\  {\bf 100}, 191601 (2008)
  [arXiv:0802.3318 [hep-th]].
  %%CITATION = PRLTA,100,191601;%%

%\cite{Buchel:2009tt}
\bibitem{Buchel:2009tt}
  A.~Buchel and R.~C.~Myers,
  %``Causality of Holographic Hydrodynamics,''
  JHEP {\bf 0908}, 016 (2009)
  [arXiv:0906.2922 [hep-th]].
  %%CITATION = JHEPA,0908,016;%%

%\cite{Hofman:2009ug}
\bibitem{Hofman:2009ug}
  D.~M.~Hofman,
  %``Higher Derivative Gravity, Causality and Positivity of Energy in a UV
  %complete QFT,''
  Nucl.\ Phys.\  B {\bf 823}, 174 (2009)
  [arXiv:0907.1625 [hep-th]].
  %%CITATION = NUPHA,B823,174;%%

%\cite{deBoer:2009pn}
\bibitem{deBoer:2009pn}
  J.~de Boer, M.~Kulaxizi and A.~Parnachev,
  %``AdS_7/CFT_6, Gauss-Bonnet Gravity, and Viscosity Bound,''
  JHEP {\bf 1003}, 087 (2010)
  [arXiv:0910.5347 [hep-th]].
  %%CITATION = JHEPA,1003,087;%%

%\cite{Camanho:2009vw}
\bibitem{Camanho:2009vw}
  X.~O.~Camanho and J.~D.~Edelstein,
  %``Causality constraints in AdS/CFT from conformal collider physics and
  %Gauss-Bonnet gravity,''
  JHEP {\bf 1004}, 007 (2010)
  [arXiv:0911.3160 [hep-th]].
  %%CITATION = JHEPA,1004,007;%%

%\cite{Buchel:2009sk}
\bibitem{Buchel:2009sk}
  A.~Buchel, J.~Escobedo, R.~C.~Myers, M.~F.~Paulos, A.~Sinha and M.~Smolkin,
  %``Holographic GB gravity in arbitrary dimensions,''
  JHEP {\bf 1003}, 111 (2010)
  [arXiv:0911.4257 [hep-th]].
  %%CITATION = JHEPA,1003,111;%%

  %\cite{Cai:2001dz}
\bibitem{Cai:2001dz}
  R.~G.~Cai,
  %``Gauss-Bonnet black holes in AdS spaces,''
  Phys.\ Rev.\  D {\bf 65} (2002) 084014
  [arXiv:hep-th/0109133].
  %%CITATION = PHRVA,D65,084014;%%


\bibitem{CNO}
  M.~Cvetic, S.~Nojiri and S.~D.~Odintsov,
  %``Black hole thermodynamics and negative entropy in deSitter and
  %anti-deSitter Einstein-Gauss-Bonnet gravity,''
  Nucl.\ Phys.\  B {\bf 628}, 295 (2002)
  [arXiv:hep-th/0112045].
  %%CITATION = NUPHA,B628,295;%%

  %\cite{Balasubramanian:1999re}
\bibitem{Balasubramanian:1999re}
  V.~Balasubramanian and P.~Kraus,
  %``A stress tensor for anti-de Sitter gravity,''
  Commun.\ Math.\ Phys.\  {\bf 208} (1999) 413
  [arXiv:hep-th/9902121].
  %%CITATION = CMPHA,208,413;%%

%\cite{Myers:1987yn}
\bibitem{Myers:1987yn}
  R.~C.~Myers,
  %``HIGHER DERIVATIVE GRAVITY, SURFACE TERMS AND STRING THEORY,''
  Phys.\ Rev.\  D {\bf 36} (1987) 392.
  %%CITATION = PHRVA,D36,392;%%

%\cite{Davis:2002gn}
\bibitem{Davis:2002gn}
  S.~C.~Davis,
  %``Generalised Israel junction conditions for a Gauss-Bonnet brane world,''
  Phys.\ Rev.\  D {\bf 67} (2003) 024030
  [arXiv:hep-th/0208205].
  %%CITATION = PHRVA,D67,024030;%%

%\cite{Hartnoll:2009sz}
\bibitem{Hartnoll:2009sz}
  S.~A.~Hartnoll,
  %``Lectures on holographic methods for condensed matter physics,''
  Class.\ Quant.\ Grav.\  {\bf 26} (2009) 224002
  [arXiv:0903.3246 [hep-th]].
  %%CITATION = CQGRD,26,224002;%%

%\cite{Gibbons:1976ue}
\bibitem{Gibbons:1976ue}
  G.~W.~Gibbons and S.~W.~Hawking,
  %``Action Integrals And Partition Functions In Quantum Gravity,''
  Phys.\ Rev.\  D {\bf 15} (1977) 2752.
  %%CITATION = PHRVA,D15,2752;%%

  %\cite{Kofinas:2006hr}
\bibitem{Kofinas:2006hr}
  G.~Kofinas and R.~Olea,
  %``Vacuum energy in Einstein-Gauss-Bonnet AdS gravity,''
  Phys.\ Rev.\  D {\bf 74} (2006) 084035
  [arXiv:hep-th/0606253].
  %%CITATION = PHRVA,D74,084035;%%


  %\cite{Brihaye:2008kh}
\bibitem{Brihaye:2008kh}
  Y.~Brihaye and E.~Radu,
  %``Five-dimensional rotating black holes in Einstein-Gauss-Bonnet theory,''
  Phys.\ Lett.\  B {\bf 661} (2008) 167
  [arXiv:0801.1021 [hep-th]].
  %%CITATION = PHLTA,B661,167;%%


%\cite{Astefanesei:2008wz}
\bibitem{Astefanesei:2008wz}
  D.~Astefanesei, N.~Banerjee and S.~Dutta,
  %``(Un)attractor black holes in higher derivative AdS gravity,''
  JHEP {\bf 0811} (2008) 070
  [arXiv:0806.1334 [hep-th]].
  %%CITATION = JHEPA,0811,070;%%

%\cite{Brihaye:2008xu}
\bibitem{Brihaye:2008xu}
  Y.~Brihaye and E.~Radu,
  %``Black objects in the Einstein-Gauss-Bonnet theory with negative
  %cosmological constant and the boundary counterterm method,''
  JHEP {\bf 0809} (2008) 006
  [arXiv:0806.1396 [gr-qc]].
  %%CITATION = JHEPA,0809,006;%%



%\cite{Ge:2010aa}
\bibitem{Ge:2010aa}
  X.~H.~Ge, B.~Wang, S.~F.~Wu and G.~H.~Yang,
  %``Analytical study on holographic superconductors in external magnetic
  %field,''
  arXiv:1002.4901 [hep-th].
  %%CITATION = ARXIV:1002.4901;%%


\end{thebibliography}
\end{document}